\documentclass[12pt]{iopart}
\begin{document}

\title{
Spherically symmetric perfect fluid in area-radial coordinates}
\author{
Hideo Iguchi\dag, Tomohiro Harada\ddag, and Filipe C Mena\S
}
\address{\dag
Department of Physics, Tokyo Institute of Technology, Oh-Okayama, Meguro, Tokyo 152-8551, Japan}
\address{\dag
Laboratory of Physics, College of Science and Technology, Nihon University, \\
7-24-1, Narashino-Dai, Funabashi, Chiba 274-8501, Japan}
\address{\ddag
Astronomy Unit, School of Mathematical Sciences, Queen Mary, University
of London, Mile End Road, London E1 4NS,  UK}
\address{\S
Mathematical Institute, University of Oxford, St. Giles 24-29, Oxford OX1 3LB, UK} 
\address{\S
Departamento de Matem\'atica,
Universidade do Minho, Campus de Gualtar,
4710 Braga, Portugal}

\begin{abstract}
We study the spherically symmetric collapse of a perfect fluid using
area-radial coordinates.
We show that analytic mass functions describe a static regular centre
in these coordinates.
In this case, a central singularity can not be realized without an infinite  
discontinuity in the central density.
We construct mass functions involving fluid dynamics at the centre 
and investigate the
relationship between those and the nature of the
singularities.


\end{abstract}

\maketitle
\section{Introduction}
\label{sec:introduction}

Gravitational collapse is one of the main issues of
gravitational physics.
The nature of spacetime singularities formed
in gravitational collapse has been studied for a long time.
In these studies,
it is commonly assumed some sort of spacetime symmetry, such as
spherical symmetry.

The final fate of spherically symmetric dust collapse
has been extensively studied. In this case, a general exact solution
to Einstein's field equations is available.
It was proved that a collapsing dust ball results
in naked or covered
singularities~\cite{christodoulou1984}.
The condition for both outcomes
has been derived in terms of the initial
data~\cite{jd1993,sj1996,jj1997} and
the structure of naked singularities formed in this
system has also been studied in detail~\cite{Mena:2001dr,Nolan:2002}.

As for more realistic fluids with equations of state, such as
$p=\alpha \epsilon$ where the pressure $p$ is proportional to the
energy density $\epsilon$, several important results have been
obtained so far. There are self-similar collapse solutions which
result in naked singularity formation~\cite{op1987,op1990}. The
generic appearance of naked singularities has been strongly
suggested numerically for a very small value of
$\alpha$~\cite{harada1998,hm2001}. The critical collapse appears
at the black hole threshold for $0<\alpha \le 1$, which can be
identified with naked singularity~\cite{ec1994,nc2000}. 
In spite of these findings, a comprehensive
understanding of the endstates of gravitational collapse in this
system has not been achieved yet. This is partly due to the fact
that we do not know a general exact solution for this system,
unlike for the dust cases.
{ It should be noted that there are several related papers to this work
by Christodoulou \cite{Christo}.}

Recently, Giamb\`o {\it et al.} gave an interesting approach to
the spherically symmetric system of a perfect fluid with the
barotropic equation of state
$p=p(\epsilon)$~\cite{{Giambo:2003fd},{Giambo:2003tp}}. They
adopted the so-called area-radial coordinates and derived a
second-order quasi-linear partial differential equation (PDE) for
a mass function, which is identical to the Misner-Sharp
quasi-local mass. Due to the presence of pressure the mass
function is time dependent. A solution to this PDE gives a
solution to the whole set of Einstein's field equations. Using
mass functions which are analytical around the centre, they have
examined the spacetime singularities resulting from the
gravitational collapse. In particular, they claimed that there are
always naked singularities for that class of mass functions.


In this paper we analyse the behaviour of the metric around the regular 
centre and show that analytic mass functions imply staticity 
around the centre.   
This means that, in this case, a central singularity can not be formed 
without an infinite discontinuity in the fluid density. 
We then
present a more general analysis of the final fate of collapse in
this system by (i) studying other classes of mass functions and
(ii) investigating naked singularities formation within such
classes.

The plan of this paper is as follows. In section \ref{sec:formulation}
we briefly recall the formulation of spherically symmetric spacetimes
in area-radial
coordinates.
In section \ref{sec:behaviour} we clarify the behaviour of
metric functions during the regular evolution, which is a key
to the present problem.
We investigate 
what type of collapse can be represented by the analytic mass functions
given by Giamb\`o {\it et al.}
\cite{{Giambo:2003fd},{Giambo:2003tp}} in section \ref{sec:comment}.
In section \ref{sec:friedmann} we
investigate the Friedmann solution and construct other dynamical mass
functions.
We investigate the relationship between the mass functions and naked
singularity
formation in section \ref{sec:naked}. In section \ref{sec:conclusion}
we give concluding remarks.
We adopt the units in which $G=c=1$.

\section{Formulation}
\label{sec:formulation}
\subsection{Comoving coordinates}
The line element of a spherically symmetric spacetime in comoving coordinates
can be written as
\begin{equation}
\label{metric}
 ds^2 = - e^{2\nu} dt^2 + e^{2\lambda} dr^2
       + R^2 (d\theta^2 + \sin^2 \theta d\phi^2),
\end{equation}
where $\nu=\nu(r,t)$, $\lambda=\lambda(r,t)$
and $R=R(r,t)$ are of class $C^2$.
We consider a perfect fluid as a matter field and denote the energy density
and the isotropic pressure of the fluid by $\epsilon$ and $p$,
respectively.
Then, the
Einstein field equations can be reduced to
\begin{eqnarray}
 \label{2.2a}
 \Psi' &=& 4 \pi \epsilon R^2 R', \\
 \label{2.2b}
 \dot{\Psi} &=& -4 \pi p R^2 \dot{R},  \\
 \label{2.2c}
 \dot{R'} &=& \dot{R} \nu' + R' \dot{\lambda}, \\
 \label{2.2d}
 p' &=& -(\epsilon + p ) \nu',
\end{eqnarray}
where a dot and a prime denote derivatives with respect to $t$ and $r$,
respectively, and $\Psi=\Psi(r,t)$ is the Misner-Sharp quasi-local mass,
\begin{equation}
 \Psi(r,t) = \frac{R}{2}\left[1-R'^2e^{-2\lambda}+\dot{R}^2 e^{-2\nu} \right].
\end{equation}
The matter (number) density is given by
\begin{equation}
\label{2.6}
\rho=\frac{e^{-\lambda}}{4\pi E R^2},
\end{equation}
where $E$ is an arbitrary positive function of $r$.
Here we assume a barotropic perfect fluid with a linear equation of state
\begin{equation}
\label{2.7}
 p=\alpha \epsilon,
\end{equation}
where $\alpha>0$ is constant. This together with $p=\rho d\epsilon/d\rho-\epsilon$
implies $\epsilon=\rho^{\alpha+1}$, where the constant of proportion
was absorbed into the function $E$. Therefore from equation (\ref{2.2d}) we get
\begin{equation}
\label{2.8}
e^{\nu}=\rho^{-\alpha},
\end{equation}
where the scaling of $t$ is chosen so that the above equation holds.
Using the freedom in scaling of $r$,
we can set at the regular initial moment
\begin{eqnarray}
R(r,0)&=&r, \label{eq:cominit1}\\
R'(r,0)&=&1, \label{eq:cominit2}
\end{eqnarray}
if there is no apparent horizon initially.
In the following we choose this scaling.

\subsection{Area-radial coordinates}
Using area-radial coordinates $(r,R)$ we find from
equations (\ref{2.2a}), (\ref{2.2b}) and (\ref{2.7})
that both $R'$ and the energy density are related to the mass function
as follows
\begin{eqnarray}
\label{2.9}
R' &=& - \frac{\alpha}{\alpha + 1} \frac{\Psi_{,r}}{\Psi_{,R}} \label{R'},\\
\label{2.10}
\epsilon &=& \rho^{\alpha+1} = -\frac{\Psi_{,R}}{4\pi \alpha R^2},
 \label{density}
\end{eqnarray}
where a subscript denotes the
partial derivative with respect to the indicated variable
in the area-radial coordinates $(r,R)$.
In these coordinates, the line element (\ref{metric})
can be written as
\begin{eqnarray}
 ds^2 &=& - \frac{1}{u^2} \left[dR^2 - 2 R' dR dr
          + \left(\frac{R'}{Y}\right)^2 \left(1- \frac{2\Psi}{R}\right)dr^2
           \right] \nonumber \\&&+ R^2(d\theta^2 + \sin^2 \theta d\phi^2),
\end{eqnarray}
where
\begin{eqnarray}
 Y &=& R' e^{-\lambda}, \label{Y_diff}\\
 u^2 &=& \frac{2\Psi}{R} + Y^2 -1. \label{u2}
\end{eqnarray}
Using equations (\ref{2.6}), (\ref{2.9}) and (\ref{2.10}), the function $Y$
can also be written in terms of
the mass function as
\begin{equation}
\label{2.13}
 Y = \frac{E \Psi_{,r}}{(\alpha + 1)\rho^{\alpha}}.
\end{equation}
We can now use equations (\ref{2.8}), (\ref{2.9}),
(\ref{2.10}) and (\ref{2.13}) in equation (\ref{2.2c})
to reduce the Einstein's field equations (\ref{2.2a})-(\ref{2.2d}) to
the following PDE
\begin{eqnarray}
 &&- \frac{1}{\alpha +1 } \left( \frac{2\Psi}{R} - 1
    + \left( 1 - \alpha \right) Y^2 \right) \frac{\Psi_{,RR}}{\Psi_{,R}}
 - 2 Y^2  \frac{\Psi_{,rR}}{\Psi_{,r}}
 + \frac{\alpha + 1}{\alpha} \frac{\Psi_{,R}}{\Psi_{,r}}
   Y^2 \frac{\Psi_{,rr}}{\Psi_{,r}}  \nonumber \\
 &&-\frac{1}{\alpha +1}
   \left(\frac{2\Psi}{R} -1 +\left( 1 - \alpha \right)  Y^2 \right)
 \frac{2 \alpha}{R} - 2Y^2 \frac{\alpha}{R}
 + \frac{\Psi}{R^2} + \frac{\Psi_{,R}}{\alpha R} \nonumber\\ &&
 + \frac{\alpha + 1}{\alpha} \frac{\Psi_{,R}}{\Psi_{,r}}
 \frac{E_{,r}}{E} Y^2 = 0.
\label{Einsteineq}
\end{eqnarray}
Now, by eliminating $E(r)$ from equation (\ref{Einsteineq}), using two
arbitrary functions $\Psi_0(r) = \Psi(r,r)$ and $Y_0(r) = Y(r,r)$,
one can obtain the following PDE for $\Psi$,
\begin{eqnarray}
 &&- \frac{1}{\alpha +1 } \left( \frac{2\Psi}{R} - 1
    + \left( 1 - \alpha \right) Y^2 \right) \frac{\Psi_{,RR}}{\Psi_{,R}}
 - 2 Y^2  \frac{\Psi_{,rR}}{\Psi_{,r}}
 + \frac{\alpha + 1}{\alpha} \frac{\Psi_{,R}}{\Psi_{,r}}
   Y^2 \frac{\Psi_{,rr}}{\Psi_{,r}}  \nonumber \\
 &&-\frac{2 \alpha}{\alpha +1} \left(\frac{2\Psi}{R} + 2 Y^2 -1 \right)
 \frac{1}{R} + \frac{\Psi}{R^2} + \frac{\Psi_{,R}}{\alpha R}
 \nonumber\\ &&+ \frac{\alpha + 1}{\alpha} \frac{\Psi_{,R}}{\Psi_{,r}}
 \left( \frac{Y'_0}{Y_0} - \frac{1}{\alpha +1}\frac{\Psi_0''}{\Psi'_0}
       -\frac{2\alpha}{(\alpha + 1)r} \right) Y^2 = 0,
 \label{Einstein_eq_2}
\end{eqnarray}
and
\begin{equation}
\label{Yequation}
 Y(r,R) = \frac{\Psi_{,r}(r,R)}{\Psi_{,r}(r,r)}
         \left[ \frac{\Psi_{,R}(r,r)R^2}{\Psi_{,R}(r,R)r^2}
                \right]^\frac{\alpha}{\alpha+1} Y_0(r).
\end{equation}
In the following, we call the above quasi-linear second-order
PDE Giamb\`o-Giannoni-Magli-Piccione (GGMP) equation.
Because we adopt the scaling of the comoving coordinate $r$
which is given by equations (\ref{eq:cominit1}) and (\ref{eq:cominit2}),
the mass function must also satisfy equation (\ref{2.9})
at the initial surface, i.e.
\begin{equation}
\label{initial_mass}
 \Psi_{,R}(r,r) = - \frac{\alpha}{\alpha + 1}\Psi_{,r}(r,r).
\end{equation}

\section{Behaviour of the metric tensor around the regular centre}
\label{sec:behaviour}
First we consider the comoving coordinate system.
The regularity condition at $r=0$ requires
\begin{eqnarray}
R(0,t)=0, \label{eq:comreg1}\\
R'(0,t)=e^{\lambda(0,t)}, \label{eq:comreg2}
\end{eqnarray}
during the evolution. { It should be noted that these conditions 
are used in Giamb\`{o} {\it et al} \cite{Giambo:2003fd}. Also, in the 
following analysis, it is sufficient to assume lower differentiability, 
$C^1$ class, for the functions $R(r,t)$, $R'(r,t)$, $e^{\lambda(r,t)}$ 
and $\rho(r,t)$ on 
$t= {\mbox{const}}$ slice.}

An estimate of equation (\ref{2.6})
at $t=0$ with a regular centre yields
\begin{equation}
E(r)\approx \frac{1}{4 \pi \rho(0,0)r^{2}},
\label{eq:E}
\end{equation}
where we have used (\ref{eq:cominit1}), (\ref{eq:cominit2}) and
(\ref{eq:comreg2}) and the weak equality is used with the
following meaning:
\begin{equation}
A(r)\approx B(r) \quad \Leftrightarrow \quad
\lim_{r\to 0}\frac{A(r)}{B(r)}=1.
\end{equation}
Then, substituting equation (\ref{eq:E}) into equation (\ref{2.6}),
and taking condition (\ref{eq:comreg2}) into account,
we get
\begin{equation}
\rho(r,t)\approx \rho(0,0)\frac{r^{2}}{R^{2}R'},
\label{eq:keyequation}
\end{equation}
where the both sides depend on $r$ and $t$ and the weak equality
``$\approx$'' is used with the following meaning:
\begin{equation}
A(r,t)\approx B(r,t) \quad \Leftrightarrow \quad
\lim_{r\to 0}\frac{A(r,t)}{B(r,t)}=1,
\end{equation}
i.e. the limit is taken as $t$ is fixed.
Integrating equation (\ref{eq:keyequation}) for small $r$
as $t$ is fixed, we have
\begin{equation}
R^{3}\approx \frac{\rho(0,0)}{\rho(0,t)}r^{3}+C(t),
\end{equation}
where $C(t)$ is an arbitrary function. From the regularity
condition (\ref{eq:comreg1}), we conclude that $C(t)=0$. Then, we
finally get
\begin{eqnarray}
R&\approx& \left[\frac{\rho(0,0)}{\rho(0,t)}\right]^{1/3}r
\label{eq:Rcentral}, \\
R'&\approx& \left[\frac{\rho(0,0)}{\rho(0,t)}\right]^{1/3}.
\label{eq:R'central}
\end{eqnarray}
This implies that the area radius $R$ is proportional to
the comoving coordinate $r$ in lowest order as $t$ is fixed.
It should be noted that the above discussion does not prove
uniform convergence.

Next we move on to the area-radial coordinate system.
The above condition can be translated into the area-radial coordinates.
A family of $t=\mbox{const}$ regular spacelike hypersurfaces
in the comoving coordinates
is transformed to a one-parameter family of
curves $R=R_{t}(r)$ in the $rR$ plane, which cross the origin $R=r=0$
along straight lines $R=\tau r$ with $0<\tau<\infty$.
If the fluid is static, the family of $t=\mbox{const}$
spacelike hypersurfaces is transformed to one degenerate
straight line $R=r$ in the $rR$ plane.
If we consider models in which the central density
changes monotonically with time, this slope parameter $\tau$
parameterises the family of evolution curves.
Moreover, if we focus on monotonically collapsing situations,
we can assume $0<\tau\le 1$.
The initial hypersurface $t=0$ corresponds to the
straight line $R=r$ with forty five degrees in the
$rR$ plane.
As time proceeds, the slope of evolution curves at the origin $R=r=0$
gradually decreases towards zero.
The $r$ axis corresponds to the singularity curve.
{ This means that we can approach a central singularity along a curve $R \approx r^n$ with $n>1$.}
The regularity condition (\ref{eq:comreg2})
is now translated into the area-radial coordinates
in terms of the function $Y$ as
\begin{equation}
\lim_{r\to 0}Y(r,\tau r)=1,
\label{eq:arreg}
\end{equation}
for $0<\tau< \infty $, where the limit is taken as $\tau$ is fixed.

\section{Analytic mass functions}
\label{sec:comment} In this section we would like to investigate
what type of collapse can be represented  by the mass functions
given by Giamb\`{o} et al.
\cite{{Giambo:2003fd},{Giambo:2003tp}}.
We recall that in \cite{Giambo:2003fd} it is assumed that $\Psi$ is an analytic
function with respect to $r$ and $R$ which can
be expanded as
\begin{equation}
 \Psi(r,R) = \sum_{k=0}^{\infty}~~\sum_{i+j=3+2k}\Psi_{ij} r^i R^j,
\end{equation}
where $i$ and $j$ are nonnegative integers,
and it is proved that the regularity condition (\ref{eq:arreg}) requires the
following structure for the lowest order terms:
\begin{equation}
 \label{expansion_giambo}
 \Psi(r,R) = \frac{h}{2}\left(r^3 - \frac{\alpha}{\alpha +1} R^3 \right)
             +\sum_{k=1}^{\infty}~~\sum_{i+j=3+2k}\Psi_{ij} r^i R^j.
\end{equation}
In addition, it is assumed that the coefficient $\Psi_{41}$ vanishes in the
proof of Lemma 3.5 of \cite{Giambo:2003fd}.
Using this form of the mass function, it is claimed that
the central singularity is naked
in Theorem 3.4 of \cite{Giambo:2003fd}.
In the following, however, we show that
their solution
has a static centre in the regular evolution.

Let us consider that we have regular evolution after the initial moment.
During the regular evolution, we can use the relation $R\approx \tau r$
($0<\tau<\infty$)
for small $r$, as discussed in section \ref{sec:behaviour}. Then,
it can be shown that the central density is
\begin{equation}
\label{centralrho} \lim_{r\to 0}\rho^{(1+\alpha)}(r,\tau
r)=\frac{3h}{8\pi(\alpha+1)},
\end{equation}
by substituting equation (\ref{expansion_giambo})
into equation (\ref{density}), that is, the central density does not
depend on $\tau$.
This guarantees, through equation (\ref{2.8}),
that the function $e^\nu$
is constant with time at the centre.
This also guarantees, through equations (\ref{eq:Rcentral})
and (\ref{eq:R'central}), that
$R\approx r$ and $R'\approx 1$.
We can also obtain the same result from
equation (\ref{R'}).
Equation (\ref{R'}) with the relation $R=\tau r$ ($0<\tau <\infty$)
implies
\begin{equation}
R'\approx \left(\frac{r}{R}\right)^{2}.
\end{equation}
Since this relation is satisfied at any time $t$,
this can be integrated as $t$ is fixed.
The result is
\begin{equation}
R^{3}(r,t)\approx r^{3}+D(t),
\end{equation}
where $D$ is an arbitrary function of $t$. The regularity
condition (\ref{eq:comreg1}) requires $D(t)=0$ and therefore
$R\approx r$ and $R'\approx 1$. Hence the regularity condition
(\ref{eq:comreg2}) implies $e^{\lambda}\approx 1$. 
Therefore, the symmetric centre can not become singular without 
an infinite discontinuity in $\rho$.

To construct mass functions which can describe genuine dynamical
situations, we shall include terms of negative power of $R$ in an
expansion of the mass function. In the next section we shall
construct such mass functions.

\section{Mass functions involving fluid dynamics}
\label{sec:friedmann}

We have pointed out some problems associated to the choice of mass functions (\ref{expansion_giambo}).
In this section, we construct other forms for $\Psi$ which can be used to study models of
gravitational collapse as well as cosmological models.

To begin with, we consider the Friedmann solution as the simplest
solution of a dynamical fluid.
In this case, the matter density is homogeneous and satisfies
\begin{equation}
 \rho \propto a(t)^{-3} = \left( \frac{R}{r} \right)^{-3},
\end{equation}
where $a$ is a scale factor. Substituting this relation into
equation (\ref{density}), we obtain,
\begin{equation}
 \label{F_density}
  \rho_0 \left( \frac{R}{r} \right)^{-3}
 = \left( - \frac{\Psi_{,R}}{4\pi \alpha R^2}\right)^{\frac{1}{\alpha +1}},
\end{equation}
where $\rho_0$ is a constant.
After integrating equation (\ref{F_density}) with respect to $R$, the mass function
can be written as
\begin{equation}
\label{fried}
 \Psi(r,R) = \frac{4 \pi \rho_0^{\alpha +1}}{3}
            \frac{r^{3(\alpha +1)}}{R^{3\alpha}}
            + f(r),
\end{equation}
where $f(r)$ is an arbitrary function. From the assumption
that $R=r$ is the initial regular slice,
we get that $R=\tau r$ is also a regular slice,
where $0 < \tau < 1$ is a constant.
When we adopt the regularity condition for $\Psi$,
we find $f(0)=0$.
Furthermore, from equation (\ref{initial_mass}), we obtain $f'=0$ and then
\begin{equation}
 f(r) = 0.
\end{equation}
Therefore, the mass function (\ref{fried}) of the Friedmann solution results in
\begin{equation}
 \label{mass_Friedmann}
 \Psi(r,R) = \Psi_3 \frac{r^{3(\alpha+1)}}{R^{3\alpha}},
\end{equation}
where $\Psi_3 =  4 \pi \rho_0^{\alpha+1}/3 $.

In order to recover the familiar form of the Friedmann solution, we note that
the metric functions in comoving coordinates $e^\nu$ and $e^\lambda$
can be written in terms of the scale factor $a$ using equation (\ref{Y_diff}),
$Y=Y_0(r)$ and $e^\nu \propto \rho^{-\alpha}$.
Then, after an appropriate coordinate transformation we get
\begin{equation}
 ds^2 = -d\tilde{t}^2 + a^2(\tilde{t})
       \left( \frac{dr^2}{Y_0^2(r)}
      + r^2 (d\theta^2 + \sin^2 \theta d\phi^2)\right).
\end{equation}

Next, we would like to investigate more general situations.
A physically reasonable mass function should satisfy not only the PDE
(\ref{Einstein_eq_2}) but also relation (\ref{initial_mass})
on the initial slice
as well as the regularity conditions at the regular centre on the regular slice
$R=\tau r$. In addition to these conditions, we can impose condition
(A) that the mass function is the same as in the Friedmann solution (\ref{mass_Friedmann})
at the lowest order around the symmetric centre.
To satisfy condition (A), the generalised mass function is written as
\begin{equation}
  \label{mass_general}
  \Psi(r,R) = \Psi_3 \frac{r^{3(\alpha+1)}}{R^{3\alpha}}
              +\Pi(r,R),
\end{equation}
where $\Pi(r,R)$ is a higher-order term on the regular slice, related to the inhomogeneity of mass distribution.
The functional form of $\Pi(r,R)$ is important for the final fate of
the collapse, that is, whether it is naked or covered.

In addition to the above restrictions on the mass function, we may
impose condition (B) that the mass function recovers an
appropriate dust limit as $\alpha\to 0$, in which the mass
function depends only on the comoving radial coordinate $r$. We
can consider that the mass function $\Psi$ is given by the
following series:
\begin{equation}
\label{psiser}
\Psi=\sum_{k=1}^{\infty}\Psi_{2k+1}r^{2k+1}
\left(\frac{R}{r}\right)^{\beta_{2k+1}},
\label{eq:simple_expansion}
\end{equation}
where $\{\Psi_{2k+1};k=1,2,\cdots\}$
are constant coefficients and $\{\beta_{2k+1};k=1,2,\cdots\}$
are constants which depend only on $k$ and $\alpha$.
Equation (\ref{initial_mass}) yields $\beta_{2k+1}=-(2k+1)\alpha$.
This satisfies both conditions (A) and (B). From this form of expansion,
we can conclude that
$\Psi=\Psi(r^{\alpha+1}/R^{\alpha})$.
This together with equation (\ref{2.9}) yields $R=\tau(t) r$,
where $\tau$ is an arbitrary function of $t$.
This implies that the singularity is simultaneous and therefore covered.

A potentially more interesting form for the mass function, results from a slight generalisation of
(\ref{psiser}):
\begin{equation}
\label{psigen}
 \Psi=\sum_{k=1}^{\infty}\sum_{i=1}^{\infty}
\Psi_{2k+1,i} r^{2k+1}\left(\frac{R}{r}\right)^{\beta_{2k+1,i}},
\label{eq:complex_model}
\end{equation}
where ${\bf \Psi}=\{\Psi_{2k+1,i};k=1,2,\cdots,i=1,2,\cdots\}$
is a constant coefficient matrix and
$\{\beta_{2k+1,i};k=1,2,\cdots,i=1,2,\cdots\}$
are constants which depend only on $\alpha$, $k$ and $i$.
Condition (A) yields
$\beta_{31}=-3\alpha$ and $\beta_{3i}=0$ for $i\ge 2$.
Condition (B) is satisfied if and only if $\beta_{2k+1,i}\to 0$
as $\alpha \to 0$.
Equation (\ref{initial_mass}) yields
\begin{equation}
\sum_{i=1}^{\infty}\Psi_{2k+1,i}\left[\beta_{2k+1,i}+(2k+1)\alpha\right]=0.
\end{equation}
This implies that if the coefficient matrix
${\bf \Psi}$ is regular,
$\beta_{2k+1,i}=-(2k+1)\alpha$ is obtained, which eventually
results in the same situation as the simplest model.
Therefore, if the mass function which admits the expansion
(\ref{eq:complex_model}) describes naked singularity formation,
then the coefficient matrix ${\bf \Psi}$ must be singular,
or $\det {\bf \Psi}=0$. In this case, there exists
at least a set of values for $k$ and $i$ such that
$\beta_{2k+1,i}\ne -(2k+1)\alpha$.

We note that while the emphasis here has been given to models of collapse,
the mass functions $\Psi$ can also be used in cosmological models.
In particular, mass functions (\ref{mass_Friedmann}) and (\ref{psiser}) can be associated to a
simultaneous big bang and can be used in the study of the early universe.

\section{Naked singular mass functions}
\label{sec:naked}

In this section we investigate
mass functions which involve spacetimes with naked singularities.
We follow a technique used by Mena and Nolan \cite{Mena:2001dr}
and developed by Giamb\`o {\it et al.} \cite{Giambo:2002xc}.
To show the existence of null geodesics emanating from the central
singularity, we show the existence of the region
from which future-directed outgoing null geodesics cannot get out
when they are traced back to the centre.
This region is between two curves
$R=R_1(r)$ and $R=R_2(r)$ ($0<R_{1}(r)<R_{2}(r)$)
both emanating from the central singularity.
Defining $\varphi(r,R)$ as the first-order derivative $dR/dr$ along
a future-directed outgoing radial null geodesic, i.e.
\begin{equation}
 \varphi(r,R) \equiv - \frac{\alpha}{\alpha +1} \frac{\Psi_{,r}}{\Psi_{,R}}
                     \left( 1- \frac{u}{Y}\right),
\end{equation}
the curve $R_{1}$ satisfies,
\begin{equation}
 \frac{dR_1(r)}{dr} \ge \varphi(r,R_1).
\end{equation}
while $R_{2}$ satisfies
\begin{equation}
\frac{dR_2}{dr} \le \varphi(r,R_2).
\end{equation}
In the following analysis, we assume that only one term in the expansion
of $\Pi(r,R)$ in (\ref{mass_general}), say
\begin{equation}
  \Pi_0 r^{5-m}R^m,
\label{eq:second_term}
\end{equation}
where $m$ is a constant, is important to examine the existence of
naked singularities. Under this assumption we will derive the
conditions on $\Pi_0$ and $m$ for the existence of radial null
geodesics emanating from the singularity which forms from
gravitational collapse.
As we have already mentioned, to
construct a solution we may need another term
\begin{equation}
\Pi_1 r^{5-n}R^n,
\end{equation}
which is of the same order as the term of equation (\ref{eq:second_term})
on the regular slice.

In this analysis, an important role is played by the apparent horizon
$R=R_{h}(r)$, where
\begin{equation}
 R_{h}(r) = 2 \Psi(r,R_{h}(r)).
\end{equation}
Under the above assumption, $R_{h}$ behaves around the centre as
\begin{equation}
  R_h = h r^{n_h},
\end{equation}
where the index $n_h\ge 1$ depends on the value of $m$ and $h$ is a
positive constant.
After a careful analysis, we obtain $n_h$ and $h$ in terms of
the value of $m$,
\begin{eqnarray}
 \label{nh1}
  n_h = \frac{3(\alpha +1)}{3\alpha +1}, & h=(2\Psi_3)^{\frac{1}{3\alpha +1}}
 & \mbox{for}~~~ m>-6\alpha -1, \\
 \label{nh2}
  n_h = \frac{3(\alpha +1)}{3\alpha +1}, & h=\left(\Psi_3 + \sqrt{\Psi_3^2 + 2\Pi_0}\right)^{\frac{1}{3\alpha +1}}
 & \mbox{for}~~~ m=-6\alpha -1, \\
 \label{nh3}
  n_h = \frac{5-m}{1-m}, & h=(2\Pi_0)^{\frac{1}{1-m}}
 & \mbox{for}~~~ m<-6\alpha -1.
\end{eqnarray}
It can easily be shown that
\begin{equation}
 \label{phi_Rh}
 \varphi(r,R_{h}) = 0 < \frac{dR_{h}}{dr}
        = n_h h r^{n_h-1}.
\end{equation}
Therefore we can identify $R_h$ with $R_1$.

If there is a curve $R=R_x(r)$ such that
\begin{equation}
\label{naked_cond}
\mbox{(a)}~~~
R_x(r) > R_{h}(r)~~~ \mbox{and}~~~\mbox{(b)}~~~
\frac{dR_x}{dr} \le \varphi(r,R_x(r)),
\end{equation}
we can identify $R_{x}$ as $R_{2}$ and
show the existence of radial null geodesics emanating from
the central singularity.
Here we assume that the curve $R=R_x(r)$ behaves as
\begin{equation}
 R_x = x r^{n_x}
\end{equation}
{ with $n_x> 1$, which assures that $R_{x}$ emerges out of the central 
singularity.}
Condition (a) in equation (\ref{naked_cond}) can be rewritten as
\begin{equation}
 \mbox{(a1)}~~~ n_x < n_h ~~~ \mbox{or} ~~~
 \mbox{(a2)}~~~n_x = n_h ~~~ \mbox{and} ~~~
              x > h .
\end{equation}
In the rest of this section we would like to investigate
the behaviour of $R'(r,R_x(r))$ along the line $R_x$,
\begin{eqnarray}
 \label{R'x}
 R'(r,R_x(r)) &=&
   \frac{ 3\alpha(\alpha+1) \Psi_3
         +\alpha (5-m)\Pi_0x^{3\alpha+m} r^{(3\alpha + m)n_x -3\alpha -m+2}}
        { 3\alpha(\alpha+1) \Psi_3
         -(\alpha+1)m\Pi_0x^{3\alpha+m}r^{(3\alpha + m)n_x -3\alpha -m+2}}
      \nonumber \\ && \times x r^{n_x-1}
\end{eqnarray}
and the limit, 
\begin{equation}
 \label{1-u/Y}
 \lim_{r\to 0}\left(1 - \frac{u(r,R_x(r))}{Y(r,R_x(r))}\right)= C_x
\end{equation}
where $C_x$ is a constant for each $x$.
Then, $\varphi(r,R_x(r))$ can be represented as
\begin{equation}
 \varphi(r,R_x(r)) \approx C_x R'(r,R_x(r))
\end{equation}
around $r=0$.

When the inequality
\begin{equation}
 3\alpha + m -2 < (3\alpha + m) n_x
\end{equation}
holds, the first terms of the denominator and the numerator are
leading terms and then the right-hand side of equation (\ref{R'x})
is approximated as  $R'(r,R_x) \approx x r^{n_x -1}$.
{ Also, it can be shown that $Y(r,R_x(r)) \approx 1$ around the centre.
Therefore,}
outside the apparent horizon, it can be easily shown that 
the limit $C_x$ is less than or equals $1$.
Therefore the following relation holds
\begin{equation}
  \varphi(r,R_x(r)) \approx C_x x r^{n_x -1} 
     < n_x x r^{n_x -1}= \frac{dR_x}{dr}.
\end{equation}
As a result, condition (b) in equation (\ref{naked_cond}) cannot be
satisfied and the central singularity would not be naked in this
case.

Next we consider the case
\begin{equation}
 \label{1=2}
 3\alpha + m -2 = (3\alpha + m) n_x,
\end{equation}
where $R'(r,R_x)$ behaves as
\begin{equation}
  R'(r,R_x) \approx
     \frac{ 3\alpha(\alpha+1) \Psi_3
         +\alpha (5-m)\Pi_0x^{3\alpha+m}}
        { 3\alpha(\alpha+1) \Psi_3
         -(\alpha+1)m\Pi_0x^{3\alpha+m}}
      x r^{n_x-1},
\end{equation}
around the centre.
Also, the following relations hold around the centre,
\begin{equation}
 Y (r,R_x(r)) \approx 
    \frac{1 + \frac{(5-m)\Pi_0}{3(\alpha + 1) \Psi_3}x^{3\alpha+m}}
    {\left(1 - \frac{m \Pi_0}{3\alpha \Psi_3} x^{3\alpha+ m} 
                                       \right)^{\frac{\alpha}{\alpha+1}}}
\end{equation}
\begin{equation}
 \frac{2\Psi(r,R_x(r))}{R_x(r)} \approx \left(
\frac{2\Psi_3}{x^{3\alpha+1}} + 2 \Pi_0 x^{m-1} \right) r^{3(\alpha +1) -(3\alpha +1)n_x}.
\end{equation}
When $m > -6\alpha -1$, condition (a) means
\begin{equation}
 n_x \le 1 + \frac{2}{3\alpha +1}.
\end{equation}
Using this inequality and equation (\ref{1=2}), it can be shown that
\begin{equation}
 m = -3 \alpha - \frac{2}{n_x -1} \le -6\alpha -1,
\end{equation}
which is inconsistent with  $m > -6\alpha -1$.

When  $m = -6\alpha -1$, we obtain from equation (\ref{1=2})
\begin{equation}
 n_x = \frac{3(\alpha + 1)}{3\alpha+1}= n_h.
\end{equation}
To satisfy condition (a2) the following relation should hold
\begin{equation}
 \label{a2_2}
 x > h=\left(\Psi_3 + \sqrt{\Psi_3^2 + 2\Pi_0}\right)^{\frac{1}{3\alpha +1}}.
\end{equation}
The  condition (b) can be rewritten in
\begin{eqnarray}
\hspace{-2.cm}
\frac{3(\alpha +1)}{3\alpha +1} &\le&
\left\{ 1-\sqrt{1 - 
           \frac{\left( 1 - 2 \Psi_3 x^{-3\alpha-1} - 2\Pi_0 x^{-6\alpha-2}
                  \right)
          \left(1 + \frac{(6\alpha+1) \Pi_0}{3\alpha \Psi_3} x^{-3\alpha-1} 
                                       \right)^{\frac{2\alpha}{\alpha+1}}}
              {\left(1 + \frac{2\Pi_0}{\Psi_3}x^{-3\alpha-1}
               \right)^2}}
    \right\} \nonumber\\
 &\times& \frac{ 1 + \frac{2\Pi_0}{\Psi_3}x^{-3\alpha-1}}
        { 1 +\frac{(6\alpha+1) \Pi_0}{3\alpha \Psi_3}x^{-3\alpha-1}} .
 \label{b_2a}
\end{eqnarray}
To satisfy this inequality, the right hand side of it should be larger than 
unity. Therefore we need
\begin{equation}
\frac{ 1 + \frac{2\Pi_0}{\Psi_3}x^{-3\alpha-1}}
        { 1 +\frac{(6\alpha+1) \Pi_0}{3\alpha \Psi_3}x^{-3\alpha-1}}
   > 1
\end{equation}
at least. This inequality holds when $\Pi_0<0$ and
\begin{equation}
 \label{x3alpha1}
 x^{3\alpha+1} > - \frac{(6\alpha+1) \Pi_0}{3\alpha \Psi_3}.
\end{equation}
We note that the positive matter density condition $\Psi_{,R}<0$ and the
non-shell crossing condition $\Psi_{,r}(r,R)>0$, which are derived from
equations (\ref{density}) and (\ref{R'}),
are satisfied in these cases.
Now we can rewrite equation (\ref{b_2a}) as
\begin{eqnarray}
\hspace{-2cm} 
&& - \left( 1 - 2 \Psi_3 x^{-3\alpha-1} - 2\Pi_0 x^{-6\alpha-2}
                  \right)
          \left(1 + \frac{(6\alpha+1) \Pi_0}{3\alpha \Psi_3} x^{-3\alpha-1} 
                                       \right)^{\frac{\alpha-1}{\alpha+1}}
\nonumber\\
\hspace{-1.5cm} &\le& 
-2 \frac{3(\alpha +1)}{3\alpha +1}\left(1 + \frac{2\Pi_0}{\Psi_3}x^{-3\alpha-1}
               \right)
+ \left(\frac{3(\alpha +1)}{3\alpha +1}\right)^2\left(1 + \frac{(6\alpha+1) 
\Pi_0}{3\alpha \Psi_3} x^{-3\alpha-1} 
               \right).
\end{eqnarray}
This inequality can hold, for example,  
if  $\alpha <1$ and $x^{3 \alpha + 1}$ 
is sufficiently near its lower limit 
$ - \frac{(6\alpha+1) \Pi_0}{3\alpha \Psi_3}$ given by (\ref{x3alpha1}).
To be compatible with equation (\ref{a2_2}), it may be needed 
an additional condition for $\Pi_0$ as
\begin{equation}
 -\frac{6 \alpha (3\alpha +1)}{(6\alpha + 1)^2} < \Pi_0 < 0.
\end{equation}

When $m < -6\alpha -1$, it can be shown that condition (a1) is
satisfied but (a2) is not from equation (\ref{nh3}).  In this case
Condition (b) gives
\begin{eqnarray}
\frac{3(\alpha +1)}{3\alpha +1} &\le&
\left\{ 1-\sqrt{1 - 
           \frac{\left(1 - \frac{m \Pi_0}{3\alpha \Psi_3} x^{3\alpha+ m} 
                                       \right)^{\frac{2\alpha}{\alpha+1}}}
          {\left(1 + \frac{(5-m)\Pi_0}{3(\alpha + 1) \Psi_3}x^{3\alpha+m}
              \right)^2}}
    \right\}
   \frac{ 1 + \frac{(5-m)\Pi_0}{3(\alpha+1)\Psi_3}x^{3\alpha+m}}
        { 1 -\frac{m \Pi_0}{3\alpha \Psi_3}x^{3\alpha+m}} .
 \label{b_2b}
\end{eqnarray}
Similarly to the previous case, 
we can show the existence of the parameters set which satisfies (\ref{b_2b})
when $\alpha <1$ and $\Pi_0 < 0$. In this case, 
$\Pi_0$ does not have a lower limit.

The last remaining case to analyse is
\begin{equation}
 \label{1<2}
 3\alpha + m -2 > (3\alpha + m) n_x,
\end{equation}
where the second terms of the denominator and the numerator
of equation (\ref{R'x}) are the
leading terms. Around the centre, $R'(r,R_x)$ behaves as
\begin{equation}
  R'(r,R_x) \approx - \frac{\alpha}{\alpha + 1} \frac{5-m}{m} x r^{n_x-1}.
\end{equation}
Along the curve $R_x$, 
\begin{equation}
 Y(r,R_x(r)) \approx \frac{(5-m)\Pi_0 x^m}{3(\alpha+1)\Psi_3}
          \left(\frac{-3\alpha \Psi_3}{m\Pi_0 x^{m-3}} 
              \right)^{\frac{\alpha}{\alpha+1}}
           r^{\frac{-3\alpha +2-m + (3\alpha +m) n_x}{\alpha+1}}
      \to \infty
\end{equation}
when we approach to the centre. Therefore 
\begin{equation}
 u^2(r,R_x(r)) \approx Y^2(r,R_x(r))
\end{equation}
and we conclude $C_x = 0$ for the curve which satisfies the condition (a).
This means that the condition (b) is inconsistent with (a).

In summary, there are parameters sets of $m$, $\Psi_3$ $\Pi_0$, and $x$
which satisfy $m = -6\alpha -1$ and equations 
(\ref{a2_2}) and (\ref{b_2a}) or $m < -6\alpha -1$ and equation (\ref{b_2b}). 
In these cases, there are curves $R=R_h(r)$ and $R=R_x(r)$,
where $R_h$ satisfies equation (\ref{phi_Rh}) and
$R_x$ satisfies equation (\ref{naked_cond}).
Therefore when the mass function with such parameters
is a solution of the GGMP equation, 
{ for which there is a central singularity after a finite time, then}
we can show the existence of
radial null geodesics which emanate from central singularity.

We note that the present analysis does not fully prove that the
spherical collapse of a perfect fluid (with any $\alpha$) produces
naked singularities since in order to do that one should check,
for each case, the consistency between the expanded form of mass
function and the GGMP equation. After that we may use
the above results to obtain a restriction on $\alpha$ which
produces a naked singularity.
Another way to do this would be to investigate whether the
solution of equation (\ref{Einstein_eq_2}) admits an expansion where
the second-order term is given by equation (\ref{eq:second_term}) with
$m \le -6\alpha -1$ and the existence $x$ which satisfies equations
(\ref{a2_2}) and (\ref{b_2a}) or equation (\ref{b_2b}).
However, in general these tasks are much more difficult.
As a final remark, we also note that while using the above
procedure it should be checked, for each $\Psi$, whether the
behaviours of $R_h$ and $\varphi(r,R_x(r))$ are, in fact, determined by
the second-order term of the mass function.

\section{Concluding remarks}
\label{sec:conclusion}

In this article we have investigated the collapse of
spherically symmetric perfect fluids with the equation of state
$p=\alpha \epsilon$ using area-radial coordinates. In this
coordinate system, the physical variables are written in terms of
the mass function which obeys a second-order PDE, which we call
the GGMP equation. We have found that the assumption of
analyticity of the mass function
has the problem of implying staticity around the regular centre.
In this case, the only way to form a central singularity is to
permit an infinite jump of a physical variable.
Furthermore, in order to have genuine
dynamics, we argued that the expansion of the mass function around
the centre should have negative powers of $R$.

We have then constructed other forms for the mass function,
starting with the Friedmann solution as the simplest dynamical
solution of the GGMP equation. We have then considered classes of mass
functions which represent inhomogeneous dynamical fluids. Such
mass functions approach the Friedmann solution at the lowest order
around the symmetric centre.
We have also considered a condition for which the solution has a
meaningful dust limit. We note that the power indices and the
coefficients of the mass function expansions were not given a
priori and should be determined, in each case, by the GGMP equation.

Using generalised forms for the mass function, we have
investigated the conditions for the appearance of naked
singularities resulting from collapse.
We have proved that if (i) $\Psi$ is a solution of the
GGMP equation and is given by (\ref{psigen}) with the
second-order term proportional to $r^{5-m}R^m$ with $m \le
-6\alpha -1$, 
{ for which there is a central singularity after a finite time} 
and (ii) there are $x$ which satisfy equations
(\ref{a2_2}) and (\ref{b_2a}) or equation (\ref{b_2b})
, then there are naked singularities.
One should be cautioned, however, that this result does not fully
prove the existence of naked singularities.
The work towards a complete proof of this is in progress.

\ack We would like to thank F.~Giannoni, G.~Magli and R.~Giamb\`o
for fruitful discussions, and anonymous referees for helpful comments. 
This work was partially supported by the Grant-in-Aid for Scientific
Research (No. 11217) from the Japanese Ministry of Education, Science,
Sports, and Culture.
TH was supported by JSPS Postdoctoral Fellowship for Research Abroad. 
FCM thanks
Funda\c{c}\~ao Calouste Gulbenkian for
grant 21-58348-B and Centro de Matem\'atica, Universidade do Minho, 
for support.


%

\end{document}